# Introduction to Xgrid:
# Cluster Computing for Everyone

Barbara J. Breen[1], John F. Lindner[2]

[1]*Department of Physics, University of Portland, Portland, Oregon 97203*
[2]*Department of Physics, The College of Wooster, Wooster, Ohio 44691*

Xgrid is the first distributed computing architecture built into a desktop operating system. It allows you to run a single job across multiple computers at once. All you need is at least one Macintosh computer running Mac OS X v10.4 or later. (Mac OS X Server is *not* required.) We provide explicit instructions and example code to get you started, including examples of how to distribute your computing jobs, even if your initial cluster consists of just two old laptops in your basement.

## 1. INTRODUCTION

Apple's Xgrid technology enables you to readily convert any ad hoc collection of Macintosh computers into a low-cost supercomputing cluster. Xgrid functionality is integrated into every copy of Mac OS X v10.4. For more information, visit http://www.apple.com/macosx/features/xgrid/.

In this article, we show how to unlock this functionality. In Section 2, we guide you through setting up the cluster. In Section 3, we illustrate two simple ways to distribute jobs across the cluster: shell scripts and batch files. We don't assume you know what shell scripts, batch files, or C/C++ programs are (although you will need to learn). Instead, we supply explicit, practical examples.

## 2. SETTING UP THE CLUSTER

In a typical cluster of three or more computers (or processors), a **client** computer requests a job from the **controller** computer, which assigns an **agent** computer to perform it. However, the client, controller, and agent can be the same computer. In fact, we initially set up a "mini-grid" consisting of just two laptop computers, a G3 and a G4 PowerBook.

### *2.1. Assigning Agents*

Designating an agent is a simple procedure handled in the Mac OS SYSTEMS PREFERENCES. For any computer running Mac OS X v10.4 "Tiger", simply check the Xgrid option under the Sharing pane of SYSTEMS PREFERENCES, as in Fig. 1. (Older computers running Mac OS X v10.3 "Panther" can download a dedicated agent installer from http://www.apple.com/support/downloads/xgridagentformacosx103.html.)

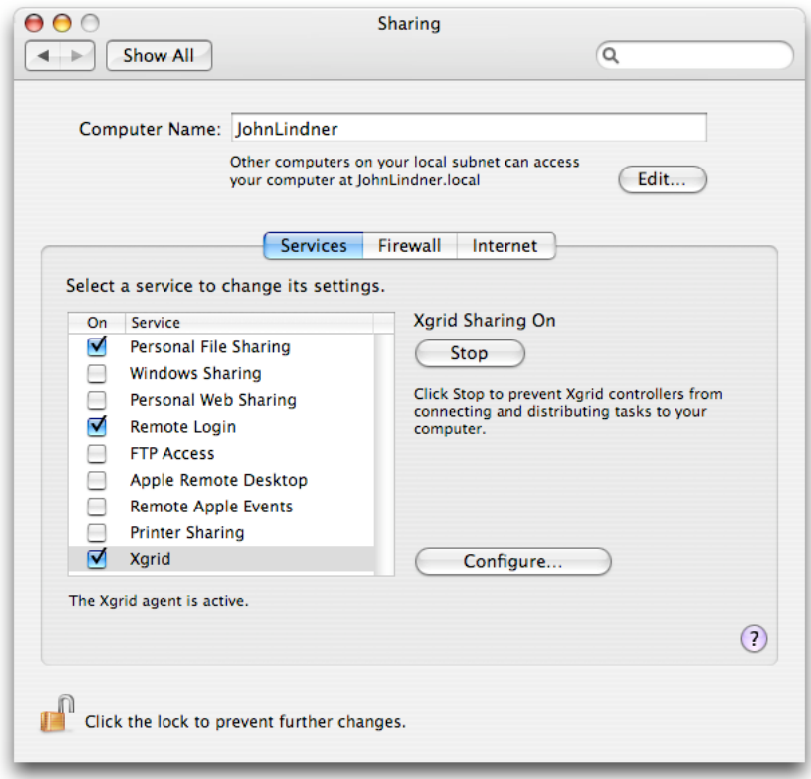

*Fig. 1. Creating an agent by checking Xgrid in the Sharing pane of System Preferences.*

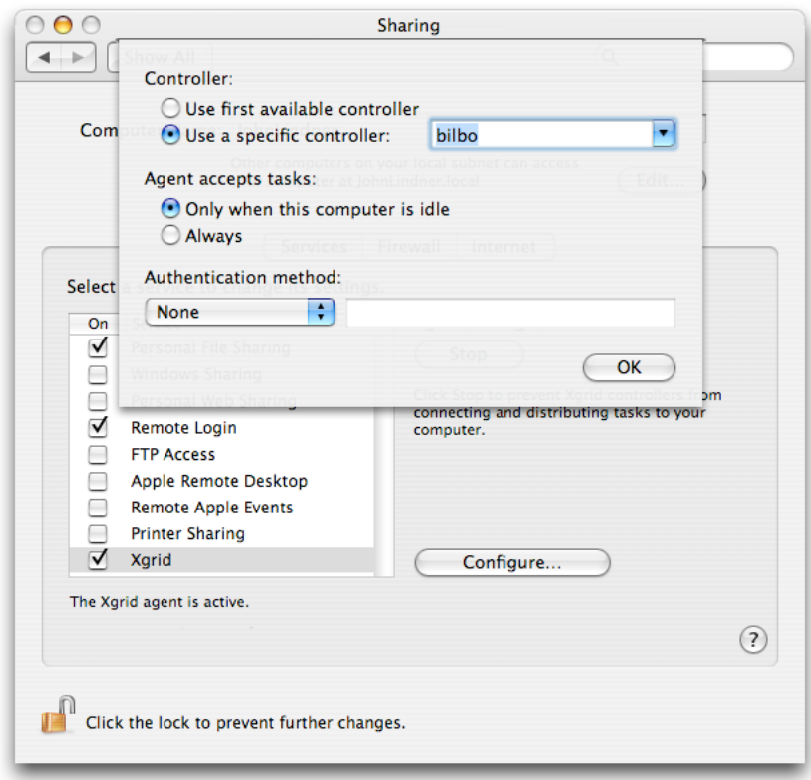

*Fig. 2. These are the simplest settings for an agent.*

Once the box is checked, the computer is available to act as an agent in a distributed network. Clicking the 'Configure' button opens a dialog box that allows this network to be described. The simplest configuration is to tell the computer to use the first available controller (though a specific controller can be specified), to always accept tasks (there is an option to accept tasks only when idle) and to not require authentication (the other options here are password or Kerberos single sign-on), as in Fig. 2.

## 2.2. Assigning a Controller

### 2.2.1. XGRIDLITE

The friendliest way to start (and stop) an Xgrid controller is to download the free XGRIDLITE from http://edbaskerville.com/software/xgridlite/. It installs an XGRIDLITE icon under "Other" in System Preferences, thereby enabling you to turn on and off controllers in a way similar to turning on and off agents.

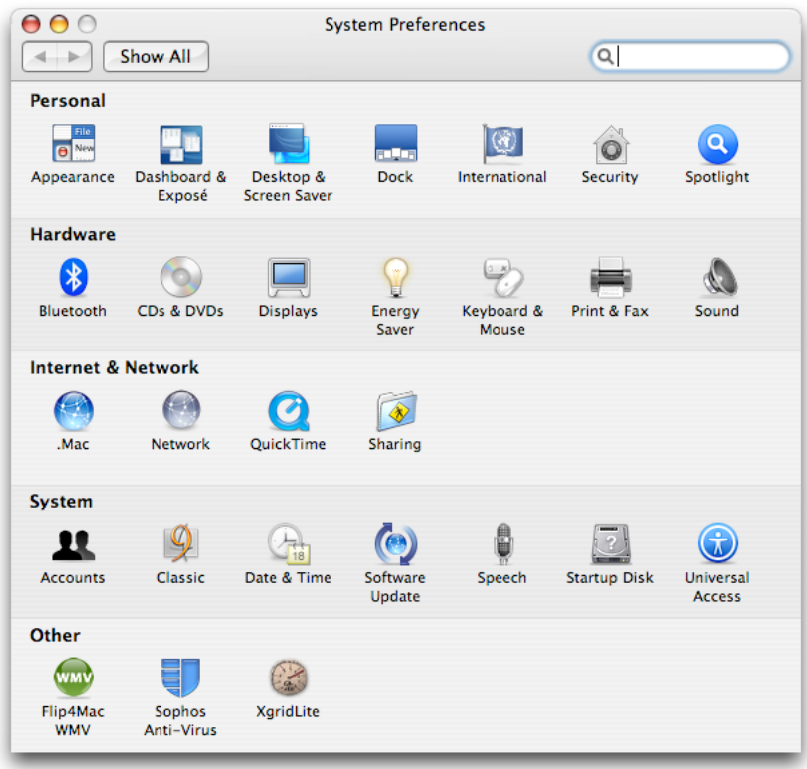

*Fig. 3. Note the XGRIDLITE icon under "other".*

Clicking on the XGRIDLITE icon opens a dialog box that lets you configure the machine as the controller. You can also choose whether or not to require authentication. (This authentication will be a non-Kerberos password; you can't enable single sign-on with XGRIDLITE.)

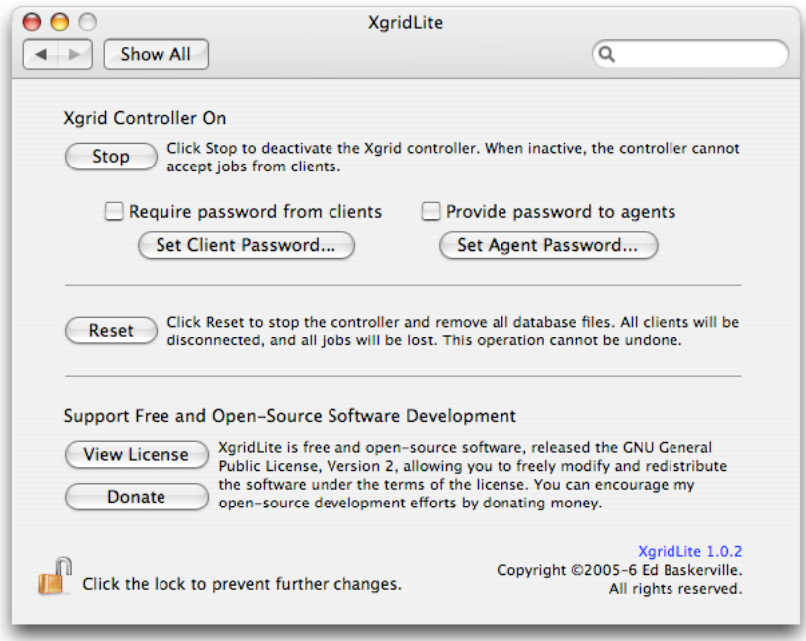

*Fig. 4. One mouse click stops or starts a controller.*

### 2.2.2. TERMINAL

Alternately, you can turn on the controller directly from the TERMINAL app using the command-line tool `xgridctl`. Simply type the command

```
breen$ sudo xgridctl c start
```

and provide an administrator's password when requested. The word `sudo` stands for "super user do" and "c" refers to the controller (rather than "a" for agent, another option). To familiarize yourself with this tool, access the manual by typing the command

```
breen$ man xgridctl
```

## 2.3. Communicating with the Cluster

The simplest way to communicate with the cluster is via the TERMINAL app. Each session (every time a TERMINAL window is opened), you should set the controller hostname and password to avoid having to continually type these at every Xgrid command. From the Mac OS X v10.4 default bash terminal, type

```
breen$ export XGRID_CONTROLLER_HOSTNAME=<hostname>
breen$ export XGRID_CONTROLLER_PASSWORD=<password>
```

From the alternate tcsh terminal, type

```
breen$ setenv XGRID_CONTROLLER_HOSTNAME <hostname>
breen$ setenv XGRID_CONTROLLER_PASSWORD <password>
```

The password doesn't have to be set if you are not requiring authentication, something we recommend in the early stages to ease the learning curve just a bit. If the computers are on the same network, you may be able to designate the controller hostname using the ".local" extension given in the SYSTEM PREFERENCES AppleTalk pane, for example "breenCluster.local". Otherwise, you should find the controller's full name including domain, such as "breenCluster.up.edu". IP addresses will also work.

Xgrid commands have the form

```
xgrid <options> <action> <parameters>
```

Immediately after setting the controller hostname, type

```
breen$ xgrid -grid list
```

If the Xgrid is functioning, you should get something like

```
{gridList = (0); }
```

meaning that there is one Xgrid of ID 0 on your network. If you don't get this response, you've probably erred in setting the hostname or password. *Watch for typos!* Any time the terminal doesn't understand a command line instruction starting with `xgrid`, it will return the `xgrid` manual page, which you can also helpfully invoke by typing

```
breen$ man xgrid
```

(Exit from the manual by typing "control-z".) You can get the attributes of an Xgrid with

```
breen$ xgrid -grid attributes -gid 0
{gridAttributes = {gridMegahertz = 0; isDefault = YES; name = Xgrid; }; }
```

You can run the Unix command "echo" on the cluster by typing

```
breen$ xgrid -job run /bin/echo "Hello, World"
Hello, World
```

You can submit the job "cal" (for calendar) to the cluster by typing

```
breen$ xgrid -job submit /usr/bin/cal 05 2007
{jobIdentifier = 80411; }
```

retrieve the results by noting the job ID number (80411 in this example) and typing

```
breen$ xgrid -job results -id 80411
      May 2007
 S  M Tu  W Th  F  S
          1  2  3  4  5
 6  7  8  9 10 11 12
13 14 15 16 17 18 19
20 21 22 23 24 25 26
27 28 29 30 31
```

and get the job's attributes via

```
breen$ xgrid -job attributes -id 80411
{
    jobAttributes = {
        activeCPUPower = 0;
        applicationIdentifier = "com.apple.xgrid.cli";
        dateNow = 2007-05-23 22:22:39 -0400;
        dateStarted = 2007-05-23 22:22:15 -0400;
        dateStopped = 2007-05-23 22:22:16 -0400;
        dateSubmitted = 2007-05-23 22:22:14 -0400;
        jobStatus = Finished;
        name = "/usr/bin/cal";
        percentDone = 100;
        taskCount = 1;
        undoneTaskCount = 0;
    };
}
```

## 2.4. Monitoring the Cluster

We recommend downloading Apple's free Mac OS X server tool, XGRID ADMIN, from http://www.apple.com/downloads/macosx/apple/macosx_updates/serveradmintools1047.html. This will allow you to monitor the controller, agents, and the jobs on the cluster.

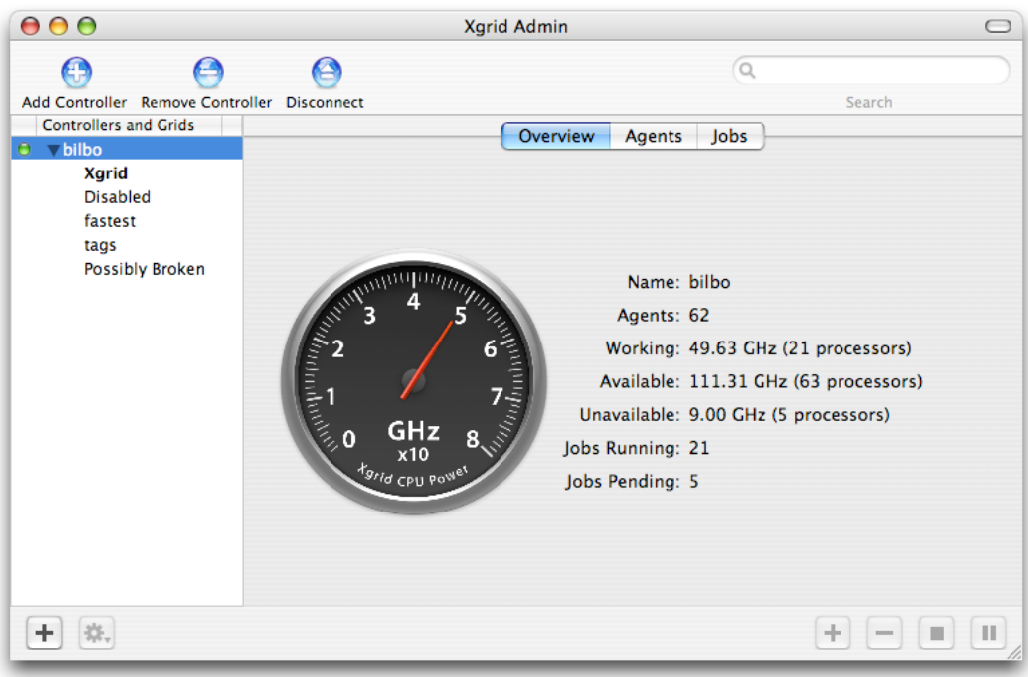

*Fig. 5. The Overview pane features a tachometer pegged at 50 GHz in this example.*

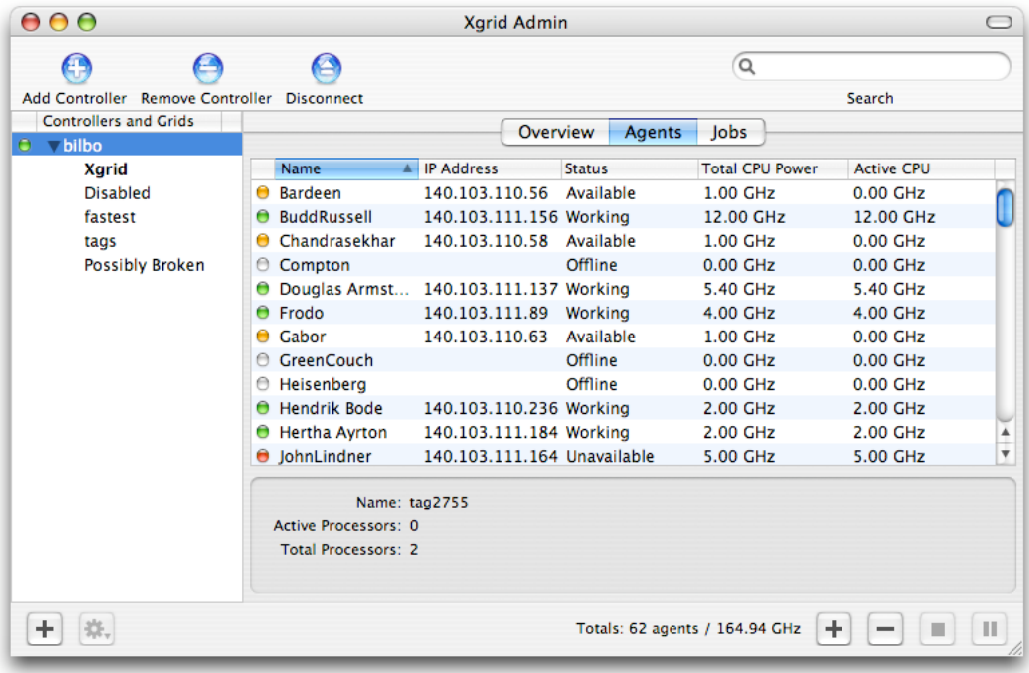

*Fig. 6. The Agents pane tells you which processors in your grid are available for jobs.*

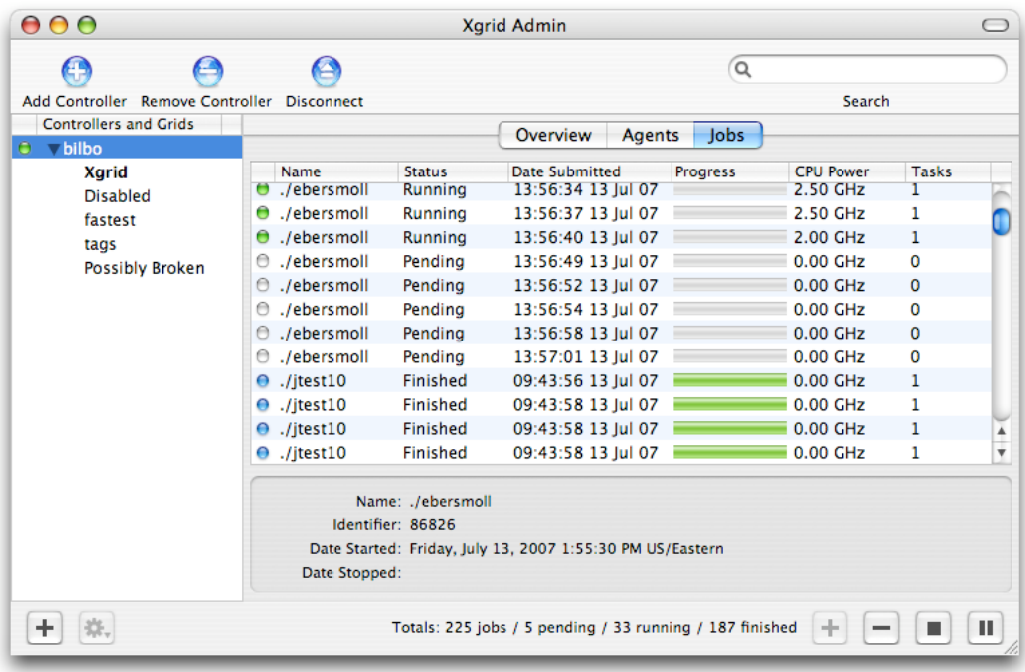

*Fig. 7. The Jobs pane allows you to monitor job progress and completion.*

# 3. USING THE CLUSTER

## 3.1. Development Environment

Once you get an appropriate response from a command line `xgrid` instruction, you are ready to test some programs. We wrote simple C/C++ programs and tested them from the command line. Then we experimented with batch jobs and shell scripts.

You can write these programs from a terminal using a classic editor like VI, or you can use an editor like TEXTWRANGLER and compile them from the command line using

```
breen$ g++ doubler.cpp -o doubler
```

However, we recommend using Apple's modern and free development tool XCODE. Launch XCODE and open "File→New Project→Command Line Utility→C++ Tool", write your code and compile using the "Build" button. The code will not execute, because it is designed to receive command line arguments, but it still compiles into an executable application that can be invoked from the command line. Copy the executable and place it in the folder from which it will be invoked from the command line.

### 3.2. C/C++ Command Line Input & Output Doubler

This sample code reads one parameter from the command line, rather than from an input file, doubles it, and outputs the result to the command line.

```cpp
//*****************doubler.cpp*****************************//
#include <stdio.h>        // for printf() etc.
#include <stdlib.h>       // for strtod()

int main (int argc, char * const argv[])
{
    if(argc - 1 != 1)                    // need 2 - 1 = 1 argument
    {
        printf("Usage: %s <number>\n", argv[0]);
        return 1;
    }
//  printf("input is %s\n", argv[1]);    // string to decimal conversion
//  double input = strtod(argv[1], NULL);   // (technique 2)

    double input;                        // string to decimal conversion
    sscanf(argv[1], "%lf", &input);      // (technique 1)

    double output = 2*input;

    printf("double %.3lf is %.3lf\n", input, output);

    return 0;
}
/************* TYPICAL TERMINAL USAGE
 *
 * breen$ ./doubler 3
 * double 3.000 is 6.000
 *
 *************/
```

### 3.3. A Job Run Shell Script

This can also be executed in Xgrid using a shell script. The run job command (unlike the submit job command) does not assign job IDs, and results are returned as the grid calculates them. After typing the code below into a text file, make it executable from the command line using

```
breen$ chmod u+rwx doublerXGrid.sh
```

```
############### doublerXGrid.sh #######################
#!/bin/sh

n=0
while [ $n -lt $1 ] ; do
    xgrid -job run ./doubler $n &
    n=$[$n + 1]
done

############### TYPICAL BASH TERMINAL USAGE #######################
#
# breen$ export XGRID_CONTROLLER_HOSTNAME=BreenLaptop.local
# breen$ xgrid -grid list
# {gridList = (0); }
# breen$ cd /Users/breen/Research/XgridCluster/XGridParamIn/
# breen$ ls -l
# total 56
# -rwxr-xr-x   1 breen  breen  17884 Dec 24 19:56 doubler
# -rw-r--r--   1 breen  wheel    964 Dec 24 20:25 doubler.cpp
# -rwxr-xr-x   1 breen  breen    980 Dec 24 20:19 doublerXGrid.sh
#
# breen$ ./doublerXGrid.sh 3
# breen$ double 2.000 is 4.000
# double 1.000 is 2.000
# double 0.000 is 0.000
#
# breen$ ./doublerXGrid.sh 3
# breen$ double 1.000 is 2.000
# double 2.000 is 4.000
# double 0.000 is 0.000
#
####################################################################
```

Note the asynchronous return of the data runs, depending on whether the task was sent to the slower of the two computers in our mini-grid.

### 3.4. C/C++ File Input & Output Multiplier

This sample code reads one parameter from a file, multiplies it by a number read from the command line, and outputs the result to a file.

```cpp
//*****************multiplier.cpp*********************************//
#include <stdio.h>                // for printf() etc.

int main (int argc, char * const argv[])
{
//   printf("{%s,  %s, %s}\n", argv[1], argv[2], argv[3]);// diagnostic

    double input;
    FILE *inputFileP = fopen(argv[1], "r");      // input data
    {
        rewind(inputFileP);
        fscanf(inputFileP, "%lf", &input);
    }
    fclose(inputFileP);

    double multiplier;                    // string to decimal conversion
    sscanf(argv[2], "%lf", &multiplier);

    double output = multiplier*input;          // transform  data
//   printf("%lf = %lf*%lf\n",output, multiplier, input); // diagnostic

    FILE *outputFileP = fopen(argv[3], "w");     // output data
    {
        fprintf(outputFileP,"%lf\n", output);
    }
    fclose(outputFileP);

    return 0;
}

/************** TYPICAL TERMINAL USAGE (WITH DIAGNOSTICS)
 * breen$ ./multiplier input.txt 10 output.txt
 * {input.txt,  10, output.txt}
 * 1.270000 = 10.000000*0.127000
 *****************************************************/
```

Once this code is vetted from the command line, you can execute it using `xgrid` commands for job submission. To multiply the input number on the Xgrid by 10, type

```
breen$ xgrid -job submit ./multiplier input.txt 10 output.txt &
{jobIdentifier = 225; }
breen$ xgrid -job results –id 225 –out output/
```

The "submit" option assigns a job ID number that can be used to retrieve the results of the job. The ampersand '&' at the end of the command line forces the job to run in the background. In other words, it gives us immediate access to the command line to do other work. Without the '&' we would be waiting in real time for the job to finish before we could access the command line. The last command retrieves the file "output.txt" and puts it in the (already created) folder "output".

### 3.5. A Job Submit Shell Script

The next test step is to run the job in Xgrid via a shell script that will vary the multiplier (argv[2] in the C/C++ code).  Note the first line of the script below: "#!/bin/sh". This has to be included for the shell program to run. This "while-do" loop submits a job to Xgrid for each value of the multiplier. The "while-do" loop (as well as the "for" loop) can only increment its argument by integer amounts. We overcome this limitation below.

```sh
############### multiplierXGrid.sh ######################
#!/bin/sh

n=0
while [ $n -lt $1 ] ; do
    xgrid -job submit ./multiplier input.txt $n output.txt -in . &
    n=$[$n + 1]
done

###############################################################
#                TYPICAL BASH TERMINAL USAGE
# breen$ ./multiplierXGrid.sh 3
# {jobIdentifier = 227; }
# {jobIdentifier = 228; }
# {jobIdentifier = 229; }
# breen$ xgrid -job results -id 227 -out output/
# breen$ xgrid -job results -id 228 -out output/
# breen$ xgrid -job results -id 229 -out output/
###############################################################
```

From the command line, the computer reads in items similarly to the way that C/C++ code reads in argv[], but in the shell script these are noted by the $ symbol. The script above has two "items" on the command line call: $0 is the executable, "./multiplierXGrid.sh", and $1 is the integer "3". Inside the script, the variable n is initialized as 0, and the loop increments it by integer amounts while it is less than 3. The Xgrid call submits three jobs, multiplier = 0, multiplier = 1 and multiplier = 2. Three job IDs result, and three job retrievals ensue.

The option "–in ." isn't really necessary here. It tells the computer that the input file is in the current directory. If necessary, a longer path name could be placed here, although we recommend copying and moving your input files to the current working directory.

## 3.6. A Batch Property List

There are a several of ways to send jobs to the cluster. The above examples use shell scripts. Another way is to send it in **batches** using **XML** code or **plist** code. The XML is more verbose and the plist is more concise. A plist can be obtained in its basic form from a properly executed Xgrid job submission, by using the option of job specification:

```
breen$ xgrid -job submit ./multiplier input.txt 10 output.txt
{jobIdentifier = 225; }
breen$ xgrid -job specification –id 225 > batch.plist
```

This saves a file of the job specifications with the name "batch.plist". It describes everything that needs to be in the plist file for a single task. We open it (in XCODE) and edit it for multiple tasks. What follows is a very simplified example of this.

### 3.7. C/C++ Command Line Input, File Output Adder

This sample code reads two parameters from the command line, adds them, and outputs the result to a file.

```cpp
/*******************************adder.cpp*******************************/

#include <stdio.h>        // for printf() etc.
#include <stdlib.h>       // for strtod()

const short maxFileName = 256;

int main (int argc, char * const argv[])
{
    if(argc - 1 != 2)                          // need 3 - 1 = 2 arguments
    {
        printf("Usage: %s <number1> <number2>", argv[0]);
        return 1;
    }

    double input1, input2;                     // read inputs
    sscanf(argv[1], "%lf", &input1);
    sscanf(argv[2], "%lf", &input2);

    double output = input1 + input2;

    printf("%.3lf + %.3lf = %.3lf\n", input1, input2, output);    // diagnostic

    char outputFileName[maxFileName];
    sprintf(outputFileName,"output_%.2lf_%.2lf.txt",input1, input2);
    FILE *outputFileP = fopen(outputFileName, "w");  // output data
    {
        fprintf(outputFileP,"%.2lf\n", output);
    }
    fclose(outputFileP);

    return 0;
}
/************** COMMAND LINE XGRID USAGE ******************
 * breen$ xgrid –job submit  ./adder 3 5
 * {jobIdentifier= 371;}
 * breen$ xgrid –job specification –id 371 >adderExec.plist
 ******************************************************/
```

In the plist batch job below, the executable is sent as an input file in base64 binary. We can't do this ourselves, that's why we have to do one run as an Xgrid job submission. The binary file goes on (nearly) forever, and we can only take a screen shot of the visible window. (It's all on one line, so it doesn't really distract). Task specifications are a list that we can add to in order to send multiple jobs.

```
################ adderExec.plist #######################
{
    jobSpecification =
      {
          applicationIdentifier = "com.apple.xgrid.cli";
          inputFiles =
            {
                "./adder" =
                  {
                      fileData = <feedface 00000012 00000000 00000002 0000000c #etc
                      isExecutable = YES;
                  };
            };
          name = "./adder";
          submissionIdentifier = abc;
          taskSpecifications =
            {
                0 = {arguments = (3, 0); command = "./adder"; };
                1 = {arguments = (3, 1); command = "./adder"; };
                2 = {arguments = (3, 2); command = "./adder"; };
                3 = {arguments = (3, 3); command = "./adder"; };
            };
      };
}

/*******COMMAND LINE XGRID USAGE (with diagnostic print statement)*******
*  breen$ xgrid -job batch adderExec.plist
* {jobIdentifier = 46954; }
* breen$ xgrid -job results -id 46954 -out .
* 3.000 + 0.000 = 3.000
* 3.000 + 1.000 = 4.000
* 3.000 + 2.000 = 5.000
* 3.000 + 3.000 = 6.000
*
*********************************************************/
```

The above adderExec.plist sends four tasks to the cluster, repeatedly calling the executable "adder" for four different combinations of numbers {3,0}, {3,1}, {3,2}, and {3,3}. The call for job results delivers four text files into the current directory ("-out ."). If you have input text files, it also works best to send them in the initial part of the plist as base64 binary.

```
            {
                "./adder" =
                  {
                      fileData = <feedface 00000012 00000000 00000002 0000000c 00000678
                      isExecutable = YES;
                  };
                "Users/breen/input1.txt" = {fileData = <302e3132 37303030 0a>; };
                "Users/breen/input2.txt" = {fileData = <302e3532 33303030 0a>; };
            };
```

The snippet of code above contains two input files, each containing only one number. Because of the extra steps involved in generating the base64 version of multiple input files (each must be submitted once as an Xgrid command line job and then the specification file retrieved to get the binary), we are leaning toward using parameter lists.

There are other forms for batch jobs, but we found this format to be most stable. It really makes a difference to send the executable once as a binary file at the beginning of the plist. You then make repeated calls to the executable in the task specifications.

### 3.8. A Simpler but Erratic Batch Property List

It is possible to skip the binary and send a complete path to the input files and the executable. However, this currently works erratically, and we don't recommend it. This may be corrected in subsequent versions of Xgrid, so we wouldn't permanently cross it off the list. Here is an example of that type of plist for a program called "batchDoubler" that reads from an input text file.

```
########### batchDoubler.plist without input/executable in binary ###########
#############           unreliable at present           ################
{
    jobSpecification =
      {
        name = doublerBatch;
        taskSpecifications =
        {
          1 = {
            arguments = ("/Users/breen/Research/XgridCluster/XGridBatch3/input1.txt");
              command = "/Users/breen/Research/XgridCluster/XGridBatch3/batchDoubler";
              };
          2 = {
            arguments = ("/Users/breen/Research/XgridCluster/XGridBatch3/input2.txt");
              command = "/Users/breen/Research/XgridCluster/XGridBatch3/batchDoubler";
              };
          3 = {
            arguments = ("/Users/breen/Research/XgridCluster/XGridBatch3/input3.txt");
              command = "/Users/breen/Research/XgridCluster/XGridBatch3/batchDoubler";
              };
          4 = {
            arguments = ("/Users/breen/Research/XgridCluster/XGridBatch3/input4.txt");
              command = "/Users/breen/Research/XgridCluster/XGridBatch3/batchDoubler";
              };
        };
    };
}
```

### 3.9. A Batch XML

You can get an XML version of a plist by typing the command

```
breen$ plutil –convert xml1 batch.plist
```

This will not rename or create a new file. It rewrites batch.plist into xml, so be sure you save a copy of it first. Then immediately rename it as, for example, batch.xml. To get multiple task XML batch files to execute reliably, we again found that it was insufficient to simply supply full path names to the additional input files. It's better to send both the executable file and any input files as binary. The XML batch file requires a different kind of binary (hexadecimal) for the executable, and the only way to get that is to first construct a multiple task plist batch file (as shown in section 3.7) and then use the command line to convert it to XML.

One drawback to working with XML batch files is that the binary version of the executable prints on multiple lines. This takes up an enormous amount of space in the visible window, if you have to open them to work on them. Since we had two methods that work reliably, the plist and the shell script, we decided to pursue those for the time being.

We include one example here of an XML batch file for multiple tasks, one that uses only parameters. Again, we have decided to avoid using input files where possible. Since any input files must first be characterized in plist format before converting to XML, we will use plist batching when input files are required.

```xml
<!--   ./adder implemented by xml batch file    -->
<?xml version="1.0" encoding="UTF-8"?>
<!DOCTYPE plist PUBLIC "-//Apple Computer//DTD PLIST 1.0//EN"
"http://www.apple.com/DTDs/PropertyList-1.0.dtd">

<plist version="1.0">

    <dict>

        <key>jobSpecification</key>
        <dict>
            <key>applicationIdentifier</key>
            <string>com.apple.xgrid.cli</string>
            <key>inputFiles</key>

            <dict>
                <key>./adder</key>
                <dict>
                    <key>fileData</key>
                    <data><!--Note hexadecimal code for executable -->
                    /u36zgAAABIAAAAAAAAAgAAAAwAAAZ4AAAAhQAAAAEA
                    <!-- etc -->
                    cgBfZnVuY3B0cgBfZnVuY3B0cgBfZnVuY3B0cgAA
                    </data>
                    <key>isExecutable</key>
                    <string>YES</string>
                </dict>
            </dict>

            <key>name</key>
            <string>./adder</string>
            <key>submissionIdentifier</key>
            <string>abc</string>
            <key>taskSpecifications</key>
```

```xml
<dict>

    <key>0</key>
    <dict>
        <key>arguments</key>
        <array>
            <string>3</string>
            <string>0</string>
        </array>
        <key>command</key>
        <string>./adder</string>
    </dict>

    <key>1</key>
    <dict>
        <key>arguments</key>
        <array>
            <string>3</string>
            <string>1</string>
        </array>
        <key>command</key>
        <string>./adder</string>
    </dict>

</dict>

        </dict>
    </dict>
</plist>
```

Below is a XML code snippet showing an input files containing one number.

```xml
<key>Users/bbreen/Research/input1.txt</key>
<dict>
    <key>fileData</key>
    <data>
    MC4xMjcwMDAK
    </data>
</dict>
<key>Users/bbreen/Research/input2.txt</key>
<dict>
    <key>fileData</key>
    <data>
    MC41MjMwMDAK
    </data>
</dict>
```

### *3.10. Shell Script to Retrieve Multiple Jobs*

Here is an example of a shell script that retrieves results from multiple jobs that have been run by a shell script. The output files are deposited in the folder "output" that already exists in the current directory.

```
#!/bin/sh

n=$1
while [ $n -le $2 ] ; do
     xgrid -job results -id $n -out output/
     n=$[$n + 1]
done

################ TYPICAL BASH TERMINAL USAGE ######################
# breen$ ./multiplierXGrid.sh 0 3 1
# breen$ {jobIdentifier = 295; }
# {jobIdentifier = 296; }
# {jobIdentifier = 297; }
# breen$ ./XgridRetrieve.sh 295 297
################################################################
```

### *3.11. Shell Script to Delete Multiple Jobs*

Below is an example of a shell script that will delete multiple jobs from Xgrid after the results have been successfully retrieved. Note the alternate addition syntax.

```
#!/bin/sh
# Deletes jobs between $1 and $2

n=$1
nEnd=`expr $1 + $2`
while [ $n -lt $nEnd ] ; do
     xgrid -job delete -id $n &
     n=`expr $n + 1`
done

################ TYPICAL BASH TERMINAL USAGE ######################
# breen$ ./xgrid_JobDelete.sh 295 297
################################################
```

### *3.12. Floating Point Shell Script Loops*

There is one last detail we needed to work around before we ran a research job. We needed to be able to increment our variables by significantly less than integer amounts. This is easy in the ksh shell. In the bash shell, we used an arbitrary precision calculator language environment called "bc". Type

```
breen$ man bc
```

for details. This language is non-intuitive and ungainly, but we found a way to assign non-integer values to our variable. In our "while-do" loop we had to evaluate an expression and take advantage of the expression returning {0, 1} for {true, false} in order to drop out of the "while" loop appropriately. Below is an example.

```
######################### complexity_bc.sh #######################
#!/bin/sh

tau=$1
while [ $echo "scale=4; $tau > $2" | bc) -lt 1 ] ; do

    xgrid -job submit ./ComplexityMetric $tau &
    tau=$(echo "scale=4; $tau + $3" | bc)

done

################# TYPICAL BASH TERMINAL USAGE ######################
# breen$ ./complexity_bc.sh 0.2 2.20.2
###############################################
```

The expression in quotes "piped" to bc gets evaluated with floating point precision of "scale=4" decimal places. So tau starts as 0.2 and is incremented by 0.2 each time through the loop until tau is greater than 2.2. Because the while statement will not accept a non-integer, we exploited the fact that, in the bc language, the inequality $tau > $2 returns 0 as long as it is true and returns 1 when it is false. Hence the structure of the while condition: "while $tau > $2 less than 1".

Another technique is to pass a series of integer-incremented numbers to the variable within the C/C++ code and to divide that number by, for example, 8, so that numbers 0 to 8 passed to the program would become 0 to 1 by 1/8 increments in the executing program.

## 4. CONCLUSION

After our basic experiments on our two-laptop basement cluster, we next utilized a brand-new rack-mounted 32-processor Xserve cluster. Apple systems engineers had already configured the Apple Workgroup Cluster with a controller node and 15 agent nodes, so we did not need to designate either status. From our office desktop, using the instructions beginning in Section 2.3 in a TERMINAL window, we were immediately able to implement the algorithms outlined in this article.

We could have used APPLE REMOTE DESKTOP to log on to the head node of the cluster and work in the TERMINAL window there. (The only difference would be the location of the client.) Our experience was that this was a bit less reliable than running the research jobs from our own desktop. We did use APPLE REMOTE DESKTOP to log on to the head node of the cluster to create folders and access the RAID storage array associated with the cluster.

We hope that this introduction will help you begin your own experiments with cluster computing.

Just do it!

**Helpful Links**

Apple Xgrid
http://www.apple.com/macosx/features/xgrid/

Apple Xgrid FAQ
http://lists.apple.com/faq/pub/xgrid_users/

MacDevCenter
http://www.macdevcenter.com/pub/a/mac/2005/08/23/xgrid.html?page=1

MacResearch
http://www.macresearch.org/the_xgrid_tutorials_part_i_xgrid_basics

Stanford Xgrid
http://cmgm.stanford.edu/~cparnot/xgrid-stanford/index.html

Utah Xgrid
http://www.macos.utah.edu/documentation/administration/xgrid/xgrid_presentation.html